\documentclass[letterpaper]{article}
\usepackage{aaai}
\usepackage{times}
\usepackage{helvet}
\usepackage[dvipsnames]{xcolor}
\usepackage{hyperref}
\usepackage{subfigure}
\usepackage[ruled,vlined]{algorithm2e} 
\usepackage{amsmath}
\usepackage{multirow}
\usepackage{multicol}
\usepackage{makecell}
\usepackage{graphicx}
\usepackage{courier}
\usepackage{booktabs}
\usepackage{array}
\graphicspath{{images/}{../images/}}

\begin{document}
 
%

\title{Automated Model Design and Benchmarking of 3D Deep Learning Models for COVID-19 Detection with Chest CT Scans}


\author{Xin He,\textsuperscript{1}
Shihao Wang,\textsuperscript{1}
Xiaowen Chu ,\textsuperscript{1}\thanks{Corresponding author}
Shaohuai Shi,\textsuperscript{2}
Jiangping Tang,\textsuperscript{3}
Xin Liu,\textsuperscript{3}
Chenggang Yan,\textsuperscript{3}\\
{\bf \Large Jiyong Zhang,\textsuperscript{3}$^*$
Guiguang Ding\textsuperscript{4}}\\
\textsuperscript{1}{Department of Computer Science, Hong Kong Baptist University, Hong Kong, China}\\
\textsuperscript{2}{Department of Computer Science and Engineering, Hong Kong University of Science and Technology, Hong Kong, China}\\
\textsuperscript{3}{School of Automation, Hangzhou Dianzi University, Hang Zhou, China}\\
\textsuperscript{4}{School of Software, Tsinghua University, Beijing, China}\\
}

\maketitle
\begin{abstract}

The COVID-19 pandemic has spread globally for several months. Because its transmissibility and high pathogenicity seriously threaten people's lives, it is crucial to accurately and quickly detect COVID-19 infection. Many recent studies have shown that deep learning (DL) based solutions can help detect COVID-19 based on chest CT scans. However, most existing work focuses on 2D datasets, which may result in low quality models as the real CT scans are 3D images. Besides, the reported results span a broad spectrum on different datasets with a relatively unfair comparison. In this paper, we first use three state-of-the-art 3D models (ResNet3D101, DenseNet3D121, and MC3\_18) to establish the baseline performance on the three publicly available chest CT scan datasets. Then we propose a differentiable neural architecture search (DNAS) framework to automatically search for the 3D DL models for 3D chest CT scans classification with the Gumbel Softmax technique to improve the searching efficiency. We further exploit the Class Activation Mapping (CAM) technique on our models to provide the interpretability of the results. The experimental results show that our automatically searched models (CovidNet3D) outperform the baseline human-designed models on the three datasets with tens of times smaller model size and higher accuracy. Furthermore, the results also verify that CAM can be well applied in CovidNet3D for COVID-19 datasets to provide interpretability for medical diagnosis. Code: \url{https://github.com/HKBU-HPML/CovidNet3D}.

\end{abstract}

\section{Introduction}
The Corona Virus Disease 2019 (COVID-19), pandemic is an ongoing pandemic caused by severe acute respiratory syndrome coronavirus 2 (SARS-CoV-2). The SARS-CoV-2 virus can be easily spread among people via small droplets produced by coughing, sneezing, and talking. COVID-19 is not only easily contagious but also a severe threat to human lives. The COVID-19 infected patients usually present pneumonia-like symptoms, such as fever, dry cough and dyspnea, and gastrointestinal symptoms, followed by a severe acute respiratory infection. The usual incubation period of COVID-19 ranges from one to 14 days. Many COVID-19 patients do not even know that they have been infected without any symptoms, which would easily cause delayed treatments and lead to a sudden exacerbation of the condition. Therefore, a fast and accurate method of diagnosing COVID-19 infection is crucial.

Currently, there are two commonly used methods for COVID-19 diagnosis. One is viral testing, which uses real-time reverse transcription-prognosis chain reaction (rRT-PCR) to detect viral RNA fragments. The other is making diagnoses based on characteristic imaging features on chest X-rays or computed tomography (CT) scan images. \cite{corr_ct_pcr} conducted the effectiveness comparison between the two diagnosis methods and concluded that chest CT has a faster detection from the initial negative to positive than rRT-PCR. However, the manual process of analyzing and diagnosing based on CT images highly relies on professional knowledge and is time-consuming to analyze the features of the CT images. Therefore, many recent studies have tried to use deep learning (DL) methods to assist COVID-19 diagnosis with chest X-rays or CT scan images.

However, the reported accuracy of the existing DL-based COVID-19 detection solutions spans a broad spectrum because they were evaluated on different datasets, making it difficult to achieve a fair comparison. Besides, most studies focus on 2D CT datasets \cite{Singh2020,Ardakani2020,Alom}. However, the real CT scan is usually the 3D data. Thus it is necessary to use 3D models to classify 3D CT scan data. To this end, we use three state-of-the-art (SOTA) 3D DL models to establish the baseline performance on the three open-source 3D chest CT scan datasets:  CC-CCII\footnote{We find there are some errors and noises in the original dataset (Version 1.0). Therefore we built our version based on it.} \cite{cell_covid}, MosMedData \cite{mosmeddata} and COVID-CTset \cite{covid-ctset}. The details are shown in Table \ref{table:datasets}.

\begin{table*}[!ht]
\caption{Summary of the existing studies of DL-based methods for COVID-19 detection. NCP indicates the novel coronavirus pneumonia, Non-NCP includes CP (common pneumonia) and Normal. $\ddagger$: the number of scans. *: the number of patients.}
\centering

\scalebox{0.74}{
\begin{tabular}{|c|c|c|c|c|c|c|c|c|c|c|c|c|}
\hline
\multirow{3}{*}{\textbf{Paper}} & 
\multirow{3}{*}{\textbf{Type}} & 
\multirow{3}{*}{\textbf{\makecell{Open-\\source?}}} & \multicolumn{3}{c|}{\textbf{Dataset Statistics}} & \multicolumn{3}{c|}{\textbf{Class Statistics (\#slices)}} & \multirow{3}{*}{\textbf{\makecell{Size of\\Test Set}}} & \multirow{3}{*}{\textbf{\makecell{2D/3D\\model?}}} & \multirow{3}{*}{\textbf{\makecell{Accuracy\\(\%)}}} \\ \cline{4-9}
 &  &   & \multirow{2}{*}{\textbf{\#patients}} & \multirow{2}{*}{\textbf{\#scans}} & \multirow{2}{*}{\textbf{\#slices}} & \multirow{2}{*}{\textbf{NCP}} & \multicolumn{2}{c|}{\textbf{Non-NCP}} &  &  & \\ \cline{8-9}
 &  &  &  &   &  &  & \textbf{CP} & \textbf{Normal} &  & & \\\hline
 \cite{Ghoshal2020} & X-ray(2D) & Yes & - & - & 5,941 & 68 & 4,290 & 1,583 & 1,188 & 2D & 88.39  \\\cline{8-9}
 \cite{Zhang2020} & X-ray(2D) & Yes &  -&  -& 1,531 & 100 & \multicolumn{2}{c|}{1,431} & 764 & 2D & -  \\\cline{8-9}
 \cite{Narin2020} & X-ray(2D) & Yes & - & - & 100 & 50 & \multicolumn{2}{c|}{50} & 20 & 2D & 98.00 \\\cline{8-9}
 \cite{Singh2020} & CT(2D) & No & - & - & 133 & 68 & \multicolumn{2}{c|}{65} & 26 & 2D & 93.20 \\\cline{8-9}
 \cite{Ardakani2020} & CT(2D) & No & 194 & - & 1,020 & 510 & \multicolumn{2}{c|}{510} & 102 & 2D & 99.63  \\\cline{8-9}
 \cite{Alom} & CT(2D) & Yes & - & - & 425 & 178 & \multicolumn{2}{c|}{247} & 45 & 2D & 98.78 \\\cline{8-9}
 \cite{He2020} & CT(2D) & Yes & 143 & - & 746 & 349 & \multicolumn{2}{c|}{397} & 186 & 2D & 86.00  \\\cline{8-9}
 \cite{Mobiny2020} & CT(2D) & Yes & - & - & 746 & 349 & \multicolumn{2}{c|}{397} & 105 & 2D & 87.60 \\\cline{8-9}
 \cite{covid-ctset} & CT(3D) & Yes & 377 & 526 & 12,058 & 244$\ddagger$ & \multicolumn{2}{c|}{282$\ddagger$} & {124$\ddagger$} & 2D & - \\\cline{8-9}
 \cite{Zheng2020} & CT(3D) & No & 542 & 630 & - & 313* & \multicolumn{2}{c|}{229*} & {131*} & 3D & 90.10 \\\cline{8-9}
 \cite{li2020artificial} & CT(3D) & No & 3,322 & 4,356 & - & 1,296$\ddagger$ & 1,735$\ddagger$& 1,325$\ddagger$ & 427$\ddagger$ & 3D & - \\\cline{8-9}
 \cite{mosmeddata} & CT(3D) & Yes & 1,110 & 1,110 & 46,411 & 856$\ddagger$ & \multicolumn{2}{c|}{254$\ddagger$} & {331$\ddagger$} & 3D & - \\\cline{8-9}

 \cite{cell_covid} & CT(3D) & Yes & 2,778 & 4,356 & 444,034 & 1,578$\ddagger$ & 1,614$\ddagger$ & 1,164$\ddagger$ & 389$\ddagger$ & 3D & 92.49\\\hline
\end{tabular}
}
\label{table:dataset_review}
\end{table*}

In addition, designing a high-quality model for the specific medical image dataset is a time-consuming task and requires much expertise, which hinders the development of DL technology in the medical field. Recently, neural architecture search (NAS) has become a prevalent topic, as it can efficiently discover high-quality DL models automatically. Many studies have used the NAS technique to image classification and object detection tasks \cite{enas,darts,nasnet_zoph17,mnasnet}. In this paper, we present a differentiable neural architecture search (DNAS) method combined with the Gumbel Softmax \cite{Gumbel_Softmax} technique to search neural architectures on three chest CT datasets: Clean-CC-CCII \cite{cell_covid}, MosMedData \cite{mosmeddata}, and COVID-CTset \cite{covid-ctset}. We represent the search space by a supernet. Using the Gumbel Softmax technique, we can optimize only one subnetwork of the supernet at a time; therefore, the searching efficiency can be significantly improved, and the search stage can be finished in about 2 hours using 4 Nvidia Tesla V100 GPUs. We name the model searched by DNAS as \textbf{CovidNet3D}. The experimental results show that CovidNet3D can achieve comparable results to human-designed SOTA models with a smaller size. Furthermore,  medical diagnoses generally require interpretability of the decision, so we apply Class Activation Mapping (CAM) \cite{cam} techniques to provide interpretability for our CovidNet3D models. In summary, our contributions are summarized as follows:
\begin{itemize}
    \item We use three manually designed 3D models to establish the baseline performance on the three open-source COVID-19 chest CT scan datasets.
    \item To the best of our knowledge, we are the first to apply the NAS technique to search for 3D DL models for COVID-19 chest CT scan datasets. Our DNAS framework can efficiently discover competitive neural architectures that outperform the baseline models on the three CT datasets.
    \item We use the Class Activation Mapping (CAM) \cite{cam} algorithm to add the interpretability of our DNAS-designed models, which can help doctors quickly locate the discriminative lesion areas on the CT scan images.
    
\end{itemize}

\section{Related Work}
In recent years, DL techniques have been proven to be effective in diagnosing diseases with X-ray and CT images \cite{Litjens2017}. To enable DL techniques to be applied in helping the detection of COVID-19, an increasing number of publicly available COVID-19 datasets have been proposed, as shown in Table \ref{table:dataset_review}. 


\subsection{Publicly-available Datasets of COVID-19}
We separate the publicly available datasets into two different categories: the pre-pandemic datasets and the post-pandemic datasets which mainly differ in quality and quantity.

\subsubsection{Pre-pandemic Datasets}
In the pre-pandemic period, gathering datasets for COVID-19 is a tough job as there is no enough data for collection. Most datasets in this period were gathered from medical papers or uploaded by the public. IEEE8023 Coivd-chestxray-dataset \cite{cohen2020covid} is a dataset of COVID-19 cases with chest X-ray and CT images collected from public sources. But its quality has no guarantee since the images are not verified by medical experts. Covid-ct-dataset \cite{zhao2020covidctdataset} is another CT dataset of COVID-19, mainly composed of CT images extracted from COVID-19 research papers, and its quality is low. The dataset only contains 2D information because each patient has only one to several CT images instead of a complete 3D scan volume.

\subsubsection{Post-pandemic Datasets}
During the pandemic, the number of confirmed cases of COVID-19 has been rising rapidly, which brings many high-quality COVID-19 chest CT scan datasets, such as CC-CCII \cite{cell_covid} and COVID-CTset \cite{covid-ctset}. Some of them have annotations by doctors, e.g., COVID-19-CT-Seg-Dataset \cite{COVID-19-CT-Seg-Dataset} and MosMedData \cite{mosmeddata}. The three datasets we use in this work are all from this category.

\subsection{DL-based methods for COVID-19 detection}
Much research is conducted on CT images, but the 3D information of CT images is under-explored, such as the work by \cite{He2020,Mobiny2020,Singh2020}. These work mainly propose 2D DL models for COVID-19 detection. \cite{Ardakani2020} benchmarks ten 2D CNNs and compares their performance in classifying 2D CT images on their private dataset with 102 testing images. On the other hand, the studies in utilizing 3D CT images are relatively rare, mainly due to the lack of 3D COVID-19 CT scan datasets. \cite{li2020artificial,Zheng2020} propose 3D CNNs with their private 3D CT datasets. There are also some other studies conducted on X-ray images. For example, \cite{Narin2020} proposes three 2D DL models for COVID-19 detection. \cite{Zhang2020} introduces a deep anomaly detection model for fast and reliable screening. \cite{Ghoshal2020} investigates the estimation of uncertainty and interpretability by Bayesian CNN on the X-ray images. \cite{Alom} uses both X-ray images and CT images to do segmentation and detection. 

\subsection{Neural Architecture Search}
In recent years, NAS has created many SOTA results by automatically searching for neural architectures for many tasks \cite{automl_survey,nas_survey}. \cite{nas2016,nasnet_zoph17} first propose to use reinforcement learning (RL) to search for neural architectures and achieves comparable results to SOTA human-designed models. Since then, several types of NAS methods have been proposed, such as evolutionary 
algorithm (EA) \cite{amoebanet}, surrogate model-based optimization (SMBO) \cite{pnas_liu18}, and gradient descent (GD) based methods \cite{darts,gdas}. \cite{gdas,fbnet} combine the GD-based method and the Gumbel Softmax \cite{Gumbel_Softmax} technique to further improve the searching efficiency.

Due to the success of NAS in natural image recognition (such as ImageNet \cite{deng2009imagenet}), researchers also try to extend it to the medical datasets, such as Magnetic resonance imaging (MRI) segmentation \cite{scalableNAS}. \cite{Faes2019} uses five public datasets, MESSIDOR, OCT images, HAM 10000, Paediatric images, and CXR images, to search for and train models by Google Cloud AutoML platform. Their experimental results demonstrate that AutoML can generate competitive classifiers compared to manually designed DL models. But to the best of our knowledge, there is no study applying the NAS technique to search for 3D DL models for COVID-19 chest CT scan datasets. To this end, we exploit the NAS technique to the three open-source COVID-19 chest CT scan datasets and successfully discover high-quality 3D models that achieve comparable performance with the human-designed SOTA 3D models.



\section{Method}
\label{section:method}


In this section, we first describe our search space for 3D CT scans classification models. Then, we introduce the differentiable neural architecture search (DNAS) method combined with the Gumbel Softmax technique \cite{Gumbel_Softmax,gdas}.

\begin{figure}[!ht]
    \centering
    \includegraphics[width=0.4\textwidth, height=0.45\textwidth]{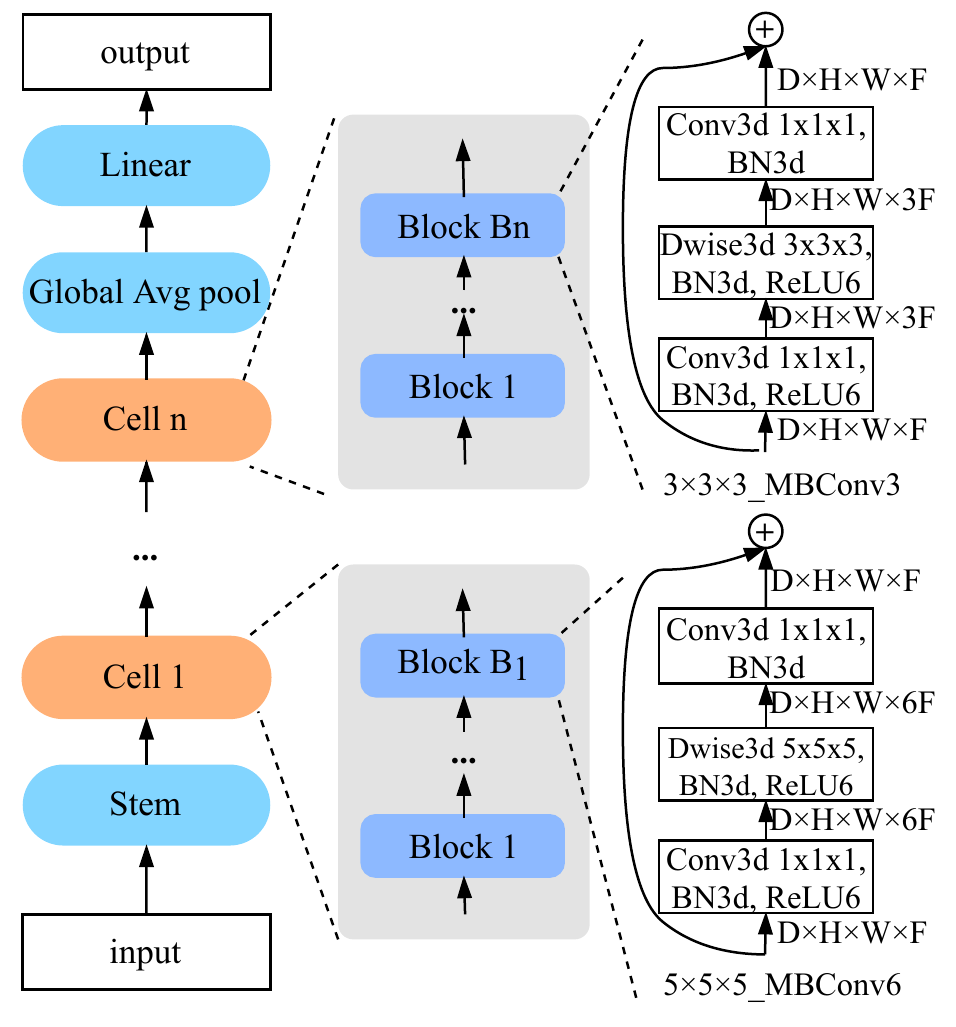}
    \caption{The overview of our search space. The model is generated by stacking a predefined number of cells. Each cell contains different number of blocks, and the block of different cells is different and needs to be searched. \textit{Conv3d 1$\times$1$\times$1} denotes 3D convolution with 1$\times$1$\times$1 kernel size, \textit{Dwise3d} denotes 3D depthwise convolution, \textit{BN3d} denotes 3D batch norm, $D\times H\times W\times F$ denotes tensor shape (depth, height, width, channel), and \textit{MBConv} denotes mobile inverted bottleneck convolution.}
    \label{fig:search_space}
    \vspace{-10px}
\end{figure}

\subsection{Search Space}

There are two critical points to be considered before designing the search space. One is that all datasets we use are composed of 3D CT scans; therefore, the searched model should be good at extracting the information from three-dimensional spatial data. The other is that the model should be lightweight, as the time required to process 3D data is much longer than 2D image data. 

Although the cell-based search space \cite{enas,darts} is one of the most commonly used search space, it has several problems: 1) the final model is built by stacking the same cells, which precludes the layer diversity; 2) many searched cells are very complicated and fragmented and are therefore inefficient for inference. MobileNetV2 \cite{mobilenetv2} is a lightweight model manually designed for mobile and embedded devices for efficient inference. Several NAS studies \cite{mnasnet,fbnet} have successfully used the layer modules \cite{mobilenetv2} including inverted residuals and linear bottlenecks to search for neural architectures and achieved SOTA results on the 2D image datasets. Therefore, we use MobileNetV2 as a reference to design our 3D search space.

As shown in Fig. \ref{fig:search_space}, we represent the search space by a supernet, which consists of the stem layer, a fixed number of cells, and a linear layer. The stem layer performs convolutional operations, and the last linear layer follows behind a 3D global average pooling operation \cite{cam}. Each cell is composed of several blocks. The structures of all blocks need to be searched. In different cells, the number of channels and the number of blocks are different and hand-picked empirically. By default, all blocks have a stride of 1. However, if a cell's input/output resolutions are different, then its first block has a stride of 2. The blocks within the same cell have the same number of input/output channels. Inspired by MobileNetV2 \cite{mobilenetv2}, each block is a MBConv-similar module (see Fig. \ref{fig:search_space}). It consists of three sub-modules: 1) a point-wise ($1\times1\times1$) convolution; 2) a 3D depthwise convolution with $K\times K\times K$ kernel size, where $K$ is a searchable parameter; 3) another point-wise ($1\times1\times1$) convolution. All convolutional operations are followed by a 3D batch normalization and a ReLU6 activation function \cite{relu6}, which is denoted by Conv3D-BN3D-ReLU6, and the last convolution has no ReLU6 activation. Another searchable parameter is the expansion ratio $e$, which controls the ratio between the size of the input bottleneck and the inner size. For example, $5\times5\times5$ \textit{MBConv6} denotes that the kernel size of \textit{MBConv} is $5\times5\times5$, and the expansion ratio is 6.

In our experiments, the search space is a fixed macro-architecture supernet consisting of 6 cells, where each has 4 blocks, but the last cell only has 1 block. We empirically collect the following set of candidate operations:

\begin{multicols}{2}
\begin{itemize}
    \item $3\times3\times3$ \textit{MBConv3}  
    \item $3\times3\times3$ \textit{MBConv4}
    \item $3\times3\times3$ \textit{MBConv6}
    \item $5\times5\times5$ \textit{MBConv3}
    \item $5\times5\times5$ \textit{MBConv4}
    \item $7\times7\times7$ \textit{MBConv3}
    \item $7\times7\times7$ \textit{MBConv4}
    \item Skip connection
\end{itemize}
\end{multicols}

Therefore, it contains $8^{21}\approx 9.2\times10^{18}$ possible architectures. Finding an optimal architecture from such a huge search space is a stupendous task. We will introduce our search strategy in the following.

\subsection{Differentiable NAS with Gumbel Softmax}


According to \cite{automl_survey}, gradient descent (GD) based NAS is an efficient method, and many studies use it to find competitive models with much shorter time and less computational resources \cite{gdas,fbnet} than other NAS methods. Hence, in this paper, we use the GD-based method and combine it with the Gumbel Softmax \cite{Gumbel_Softmax} technique to discover models for COVID-19 detection.

\subsubsection{Preliminary: DARTS}

DARTS \cite{darts} was one of the first studies to use GD-based method to search for neural architectures. Each cell is defined as a directed acyclic graph (DAG) of $N$ nodes, where each node is a network layer, and each edge between node $i$ and node $j$ indicates a candidate operation (i.e., block structure) that is selected from the predefined operation space $\mathcal{O}$. To make the search space continuous, DARTS \cite{darts} uses Softmax over all possible operations to relax the categorical choice of a particular operation, i.e.,
\begin{equation}
\begin{array}{cl}
\overline{o}_{i,j}(x)&=\sum_{k=0}^{K} P_{k} o^k(x)\\
s.t. \,\,\, P_{k} &= \frac{\exp \left(\alpha^{k}_{i,j}\right)}{\sum_{l=0}^{K} \exp \left(\alpha^{l}_{i,j}\right)}
\end{array},
\label{eq:darts_softmax}
\end{equation}
where $o^k$ indicates the $k$-th candidate operation performed on input $x$, $\alpha^k_{i,j}$ indicates the weight for the operation $o^k$ between a pair of nodes ($i,j$), and $K$ is the number of predefined candidate operations. The training and the validation loss are denoted by $\mathcal{L}_{train}$ and $\mathcal{L}_{val}$, respectively. Therefore, the task of searching for architectures is transformed into a bilevel optimization problem of neural architecture $\alpha$ and the weights $\omega_\alpha$ of the architecture:
\begin{equation}
\begin{array}{cl}
{\min_{\alpha}} & {\mathcal{L}_{val}\left(\omega_\alpha^*, \alpha\right)} \\
{\text { s.t. }} & {\omega_\alpha^*=\operatorname{argmin}_{\omega_\alpha} \, \mathcal{L}_{train}(\omega_\alpha, \alpha)}
\end{array}
\label{eq:bilevel}
\end{equation}

\subsubsection{Differentiable Model Sampling by Gumbel Softmax}

\begin{figure}[!ht]
    \centering
    \includegraphics[width=0.25\textwidth, height=0.25\textwidth]{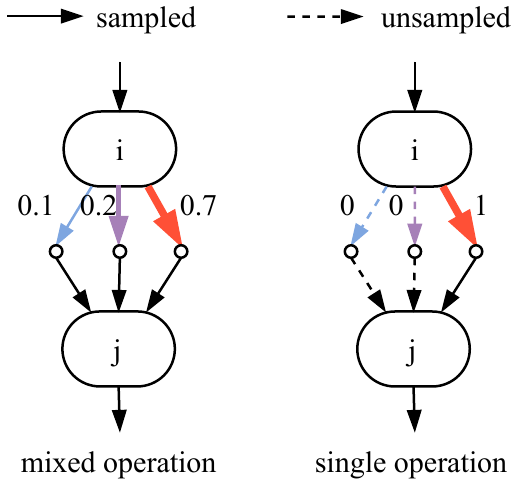}
    \caption{The comparison between two GD-based methods. (Left) Applying a mixture of all candidate operations, each with different weight. (Right) Only one operation is sampled at a time. (best viewed in color)}
    \label{fig:dag}
\end{figure}

In DARTS, as Fig. \ref{fig:dag} (left) shows, the output of each node is the weighted average of the mixed operation during the whole search stage. It causes a linear increase in the requirements of computational resources with the number of candidate operations. To alleviate this problem, we follow the same idea as \cite{gdas}. Specifically, for each layer, only one operation is sampled and executed with the sampling probability distribution $P_\alpha$ defined in Equation \ref{eq:darts_softmax}. For example, the probability of being sampled for the three operations in Fig. \ref{fig:dag} (left) is 0.1, 0.2, and 0.7, respectively, but only one operation will be sampled at a time. Therefore, the sampling distribution $P_\alpha$ of all layers is encoded into a one-hot random distribution $Z$, e.g., $P_\alpha=[0.1,0.2,0.7]\rightarrow Z=[0,0,1]$.

\begin{figure*}[!ht]
    \centering
    \includegraphics[width=0.82\textwidth, height=0.25\textwidth]{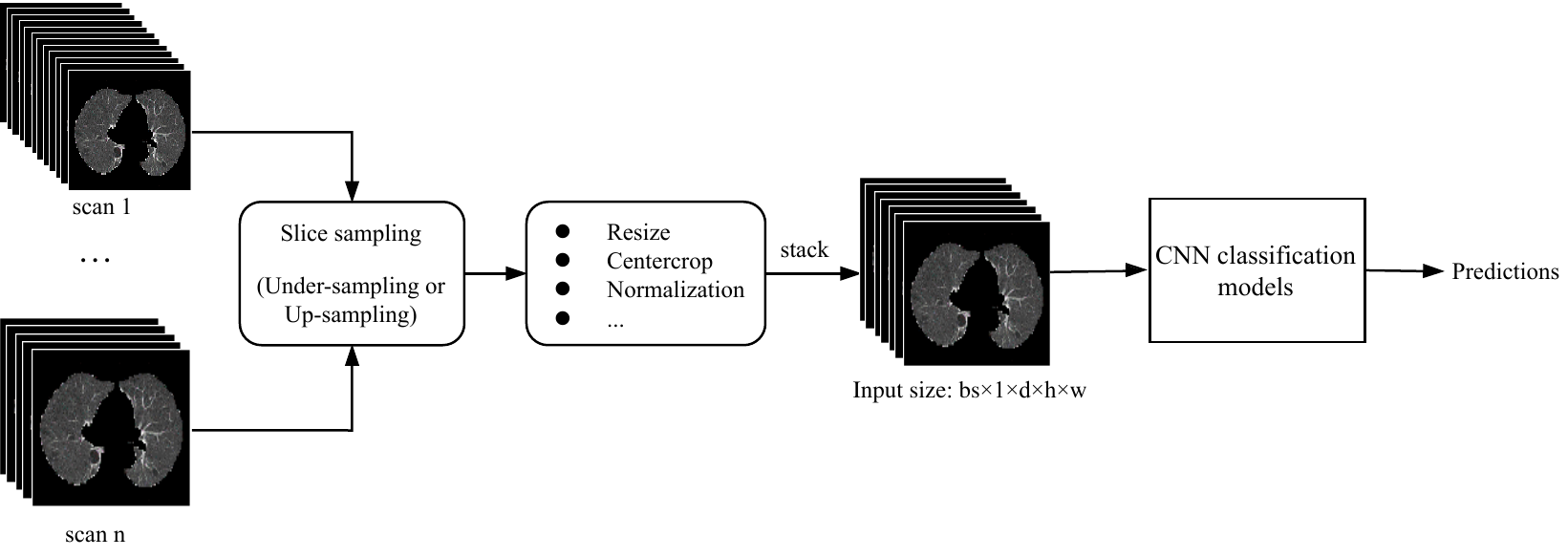}
    \caption{The pipeline of training 3D deep learning models. All CT scans need to be pre-processed by the slice sampling strategy to make sure that each scan contains the same number of slices. The input size of network is $bs\times 1\times d\times h\times w$, where $bs$ is batch size, $d$ is the number of slices, $h$ and $w$ indicate the height and width, respectively.}
    \label{fig:pipeline}
\end{figure*}

However, each operation is sampled from a discrete probability distribution $Z$, so we cannot back-propagate gradients through $Z$ to $\alpha$. To enable back-propagation, we use a reparameterization trick named Gumbel Softmax \cite{Gumbel_Softmax}, which can be formulated by
\begin{equation}
\begin{array}{cl}
Z_k=\frac{\exp \left(\left(\log {\alpha}_{i, j}^{k}+{G}_{i, j}^{k}\right) / \tau\right)}{\sum_{l=0}^{K} \exp \left(\left(\log {\alpha}_{i, j}^{l}+{G}_{i, j}^{l}\right) / \tau\right)}
\end{array},
\label{eq:gumbelsoftmax}
\end{equation}
where ${G}_{i, j}^{k}=-log(-log(u_{i,j}^k))$ is the $k$-th Gumbel sample, $u_{i,j}^k$ is a uniform random variable, and $\tau$ is the softmax temperature. When $\tau \rightarrow \infty$, the possibility distribution of all operations between each pair of nodes approximates to the one-hot distribution. To be noticed, we perform $argmax$ function on Equation \ref{eq:gumbelsoftmax} during the forward process but return the gradients according to the Equation \ref{eq:gumbelsoftmax} during the backward process.

\subsection{Class Activation Mapping Algorithm}


As mentioned above, the last linear layer follows behind a 3D global average pooling layer, which enables us to utilize class activation mapping (CAM) algorithm to generate 3D activation maps for our model. CAM exploits the global average pooling layer to calculate get the activation map $M_c$ for class $c$, where each spacial element is given by

\begin{equation}
M_{c}(x, y, z)=\sum_{k} w_{k}^{c} f_{k}(x, y, z)
\label{eq:activation_map}
\end{equation}

\noindent  where in a given image, $f_{k}(x, y, z)$ is the activation  of unit $k$ at the last convolutional layer before global average pooling layer at spatial location $(x,y,z)$, $w_{k}^{c}$ is the corresponding linear layer weight of class $c$ for unit $k$. After getting the class activation map, we can simply upsample it to the size of the input scan images to visualize and identify the regions most relevant to the specific class.

\section{Experiments}
\label{section:exp}

\subsection{Datasets and Pre-processing}
\label{section:dataset}
In this paper, we use three publicly available datasets: CC-CCII \cite{cell_covid}, MosMedData \cite{mosmeddata} and COVID-CTset \cite{covid-ctset}. The three datasets are all chest CT volumes. However, since the data format varies from the three datasets, it is necessary to pre-process each dataset to make them follow a unified way of reading data.

The original CC-CCII dataset contains a total number 617,775 slices of 6,752 CT scans from 4,154 patients, but it has five main problems (i.e., damaged data, non-unified data type, repeated and noisy slices, disordered slices, and non-segmented slices) that would have high negative impacts on the model performance. To solve these problems, we manually remove the damaged, repeated and noisy data. Then we segment the lung part for the unsegmented slice image and convert the whole dataset to PNG format. After addressing the above problems, we build a clean CC-CCII dataset named \textbf{Clean-CC-CCII}, which consists of 340,190 slices of 3,993 scans from 2,698 patients. 


\subsubsection{Scan images construction}

Each CT scan contains a different number of slices, but DL models require the same dimensional inputs. To this end, we propose two slice sampling algorithms: \textit{random sampling} and \textit{symmetrical sampling}. Specifically, the random sampling strategy is applied to the training set, which can be regarded as the data augmentation, while the symmetrical sampling strategy is performed on the test set to avoid introducing randomness into the testing results. The symmetrical sampling strategy refers to sampling from the middle to both sides at equal intervals. The relative order between slices remains the same before and after sampling. 


\begin{table}[]
\caption{The statistics of the three CT scan datasets.}
\begin{tabular}{cccccc}
\specialrule{0em}{1pt}{1pt}
\toprule
\multirow{2}{*}{\begin{tabular}[c]{@{}c@{}}Dataset\\ {[}Format{]}\end{tabular}} & \multicolumn{1}{c}{\multirow{2}{*}{Classes}} & \multicolumn{2}{c}{\#Patients} & \multicolumn{2}{c}{\#Scans} \\ \cline{3-6}  \\[-8.8pt]
 & \multicolumn{1}{c}{} & Train & Test & Train & Test \\ \hline \\[-8.8pt]
\multirow{4}{*}{\begin{tabular}[c]{@{}c@{}}Clean-\\CC-CCII\\ {[}PNG{]}\end{tabular}} & NCP & 726 & 190 & 1213 & 302 \\ \cline{2-6} \\[-8.8pt]
 & CP & 778 & 186 & 1210 & 303 \\ \cline{2-6} \\[-8.8pt]
 & Normal & 660 & 158 & 772 & 193 \\ \cline{2-6} \\[-8.8pt]
 & Total & 2164 & 534 & 3195 & 798 \\ \hline \\[-8.8pt]
\multirow{3}{*}{\begin{tabular}[c]{@{}c@{}}MosMedData\\ {[}PNG{]}\end{tabular}} & NCP & 601 & 255 & 601 & 255 \\ \cline{2-6} \\[-8.8pt]
 & Normal & 178 & 76 & 178 & 76 \\ \cline{2-6} \\[-8.8pt]
 & Total & 779 & 331 & 779 & 331 \\ \hline \\[-8pt]
\multirow{3}{*}{\begin{tabular}[c]{@{}c@{}}COVID-CTset\\ {[}16bit TIFF{]}\end{tabular}} & NCP & 80 & 15 & 202 & 42 \\ \cline{2-6} \\[-8.8pt]
 & Normal & 200 & 82 & 200 & 82 \\ \cline{2-6} \\[-8.8pt]
 & Total & 280 & 97 & 402 & 124 \\ 
\bottomrule
\end{tabular}
\label{table:datasets}
\end{table}

\subsection{Benchmarking}

We use three manually-designed 3D neural architectures as the baseline methods: DenseNet3D121 \cite{densenet3d}, ResNet3D101 \cite{videoresnet}, and MC3\_18 \cite{videoresnet}. As shown in Fig. \ref{fig:pipeline}, after building the scan images by the sampling algorithm, we further apply transformations to scans, including resize, center-crop, and normalization. Besides, for the training set, we also perform a random flip operation in the horizontal or vertical direction. The other implementation details are as follows: we use the Adam \cite{adam} optimizer and the weight decay of 5e-4. We start the learning rate of 0.001 and anneal it down to 1e-5. All baseline models are trained for 200 epochs.






\subsection{DNAS for CT Scan Classification}

\begin{figure}[!ht]
    \centering
    \includegraphics[width=0.45\textwidth]{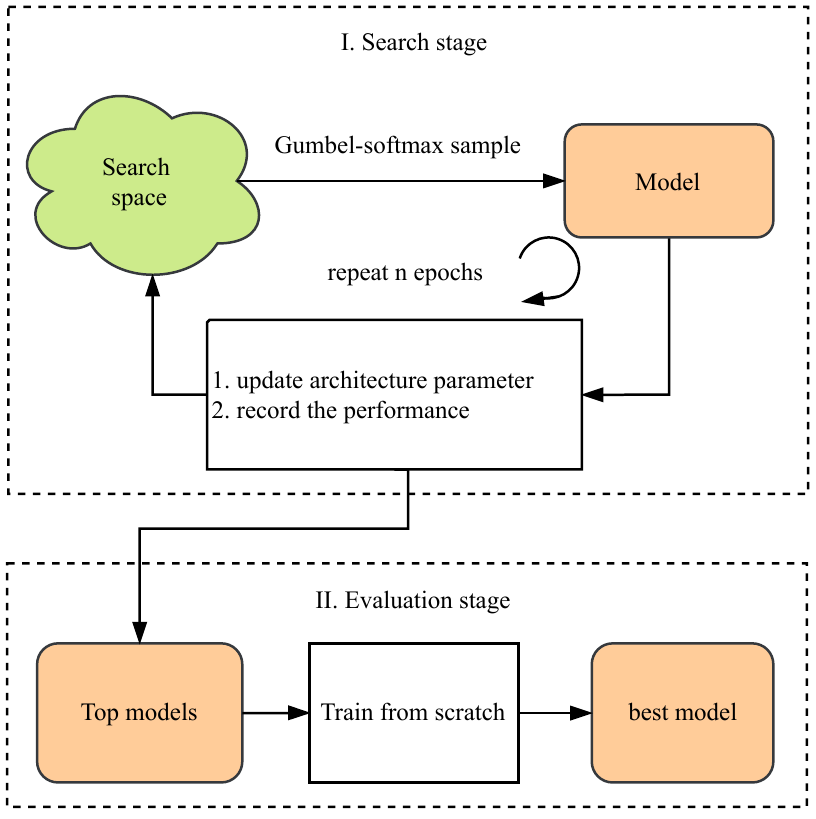}
    \caption{The pipeline of DNAS consists of two stages: the search and the evaluation stage.}
    \label{fig:gdas}
\end{figure}

To verify the efficiency of the method, we apply the DNAS method combined with the Gumbel Softmax technique to search for neural architectures on the three datasets. The pipeline of our method is shown in Fig. \ref{fig:gdas}, which contains two sequential stages: search stage and evaluation stage.

\subsubsection{Search stage}

In our experiments, the supernet consists of 6 cells with the number of blocks of $[4,4,4,4,4,1]$. Besides, the blocks within the same cell have the same number of channels. Here, we test two settings: small-scale and large-scale, where the number of channels of blocks in the 6 cells is $[24, 40, 80, 96, 192, 320]$ and $[32, 64, 128, 256, 512, 1024]$, respectively. We name the models searched under the two settings as \textbf{CovidNet3D-S} and \textbf{CovidNet3D-L}, respectively. The stem block is a Conv3D-BN3D-ReLU6 sequential module with the number of output channels fixed to 32.

To improve searching efficiency, we set the input resolution to 64$\times$64, and the number of slices in a scan to 16. We implement three independent search experiments for the three datasets. During the search stage, we split the training set into the training set $\mathcal{D_T}$ and the validation set $\mathcal{D_V}$. In each step, we first use $\mathcal{D_V}$ to update the architecture parameters $\alpha$, and then use the training set to update the sampled architecture weights $\omega_\alpha$. Besides, the architecture parameter $\alpha$ is optimized by the Adam \cite{adam} optimizer, and the architecture weights are optimized with the SGD optimizer with a momentum of 3e-4. The initial learning rate for both optimizers is 0.001.  Each experiment is conducted on four Nvidia Tesla V100 GPUs (the 32GB PCIe version) and it can be finished in about 2 hours. After each epoch, we save the sampled architecture and its performance (e.g., accuracy). Therefore, we generate 100 neural architectures for each experiment after the search stage.

\subsubsection{Evaluation stage}

As Fig. \ref{fig:gdas} shows, the search stages records the performance of the sampled architectures. In the evaluation stage, we select top-10 architectures and training these architectures with the training set for several batches, then the best-performing architecture will be retrained for 200 epochs with the full training set, and then evaluated on the test set. We set different input resolutions for three datasets to evaluate the generalization of searched architectures. Besides, since the number of slices contained in CT scans of different datasets is different, we set the intermediate value for each dataset, shown in Table \ref{table:results}. Each evaluation experiment uses the same settings as follows: we use the Adam \cite{adam} optimizer with an initial learning rate of 0.001. The cosine annealing scheduler \cite{cosineannealinglr} is applied to adjust the learning rate. We use Cross-entropy as the loss function.

\begin{table*}[!ht]
    \centering
    \caption{The experimental results of manually designed models and DNAS-designed models. }
    \scalebox{1}[1]{
    \begin{tabular}{c|cccccccc}%
    \hline
    Dataset & Model & \makecell{Model size\\(MB)} & Input size & \#Slices & \makecell{Accuracy\\(\%)} & \makecell{Precision\\(\%)} & \makecell{Sensitivity\\ (\%)} & F1-score \\ \hline
    \multirow{5}{*}{\begin{tabular}[c]{@{}c@{}}Clean-CC-CCII\\ \\ \end{tabular}}
    &ResNet3D101 & 325.21 & 128$\times$128 & 32 & 85.54 & 89.62 & 77.15 & 0.8292 \\  
    &DenseNet3D121 & 43.06 & 128$\times$128 & 32 & 87.02 & 88.97 & 82.78 & 0.8576 \\  
    &MC3\_18 & 43.84 & 128$\times$128 & 32 & 86.16 & 87.11 & 82.78 & 0.8489 \\  
    &CovidNet3D-S & 11.48 & 128$\times$128 & 32 & 88.55 & 88.78 & \textbf{91.72} & \textbf{0.9023} \\  
    &CovidNet3D-L & 53.26 & 128$\times$128 & 32 & \textbf{88.69} & \textbf{90.48} & 88.08 & 0.8926 \\ \hline
    \multirow{6}{*}{\begin{tabular}[c]{@{}c@{}}MosMedData\\ \\ \end{tabular}}
    &ResNet3D101 & 325.21 & 256$\times$256 & 40 & 81.82 & 81.31 & 97.25 & \textbf{0.8857} \\  
    &DenseNet3D121 & 43.06 & 256$\times$256 & 40 & 79.55 & \textbf{84.23} & 92.16 & 0.8801 \\  
    &MC3\_18 & 43.84 & 256$\times$256 & 40 & 80.4 & 79.43 & 98.43 & 0.8792 \\  
    &CovidNet3D-S & 12.48 & 256$\times$256 & 40 & 81.17 & 78.82 & \textbf{99.22} & 0.8785 \\  
    &CovidNet3D-L & 60.39 & 256$\times$256 & 40 & \textbf{82.29} & 79.50 & 98.82 & 0.8811 \\ \hline
     \multirow{5}{*}{\begin{tabular}[c]{@{}c@{}}Covid-CTset\\ \\ \end{tabular}}
    &ResNet3D101 & 325.21 & 512$\times$512 & 32 & 93.87 & 92.34 & \textbf{95.54} & 0.9392 \\  
    &DenseNet3D121 & 43.06 & 512$\times$512 & 32 & 91.91 & 92.57 & 92.57 & 0.9257 \\  
    &MC3\_18 & 43.84 & 512$\times$512 & 32 & 92.57 & 90.95 & 94.55 & 0.9272 \\  
    &CovidNet3D-S & 8.36 & 512$\times$512 & 32 & 94.27 & 92.68 & 90.48 & 0.9157 \\  
    &CovidNet3D-L & 62.82 & 512$\times$512 & 32 & \textbf{96.88} & \textbf{97.50} & 92.86 & \textbf{0.9512} \\ \hline
    \end{tabular}}
    \label{table:results}
\end{table*}

\section{Results and Analysis}
\label{section:results}

\subsection{Evaluation Metrics}
Our experiment results are summarized in Table \ref{table:results}. We compare our searched models with SOTA efficient models. We use several commonly used evaluation metrics to compare the model performance, as follows:
\begin{equation}
Precision=\frac{TP}{TP+FP}
\end{equation}

\begin{equation}
Sensitivity (Recall)=\frac{TP}{TP+FN}
\end{equation}

\begin{equation}
F1-score=\frac{2 \times (precision \times recall)}{precision + recall}
\end{equation}

\begin{equation}
    Accuracy= \frac{TN + TP}{TN+TP+FN+FP}
\end{equation}

To be noticed, the positive and negative cases are assigned to the COVID-19 class and the non-COVID-19 class, respectively. Specifically, $TP$ and $TN$ indicate the number of correctly classified COVID-19 and non-COVID-19 scans, respectively. $FP$ and $FN$ indicate the number of wrongly classified COVID-19 and non-COVID-19 scans, respectively. For the Clean-CC-CCII dataset, the non-COVID-19 class includes both normal and common pneumonia. The accuracy is the micro-averaging value for all test data to evaluate the overall performance of the model. Besides, we also take the model size as an evaluation metric to compare the model efficiency.


\subsection{Results on the Three CT Datasets}

Table \ref{table:results} divides the results according to the datasets. We can see that our searched CovidNet3D models outperform all baseline models on the three datasets in terms of accuracy. Specifically, CovidNet3D-L models achieve the highest accuracy of the three datasets. Besides, all CovidNet3D-S models are with much smaller sizes than the baseline models, but they can also achieve similar or even better results. For example, CovidNet3D-S (8.36 MB) achieves 94.27\% accuracy on Covid-CTset, which is 41$\times$ smaller than ResNet3D101 (325.21 MB) with 0.4\% higher accuracy. In summary, the results demonstrate that our DNAS method can discover well-performing models without inconsistency on network size, input size or scan depth (the number of slices).

We can also see that the performance of both baseline models and our CovidNet3D on the MosMedData dataset is not as good as that on the other two datasets. There are two possible reasons. One is that the MosMedData datasets's original data format is NIfTI, but all our models do not converge when trained with NIfTI files; therefore we convert NIfTI to Portable Network Graphics (PNG) format, and this process would loss information of the input files. The other possible reason is that the MosMedData dataset is imbalanced (shown in Table \ref{table:datasets}), which increases the difficulty of model training. 

We also find that the random seed greatly influences on the training of the searched CovidNet3D model through experiments. In other words, the results obtained by using different seeds for the same model would differ significantly. Hence, how to improve the robustness of NAS-based models is worthy for further exploring.

\begin{figure}[!ht]
    \centering
    \includegraphics[width=0.45\textwidth]{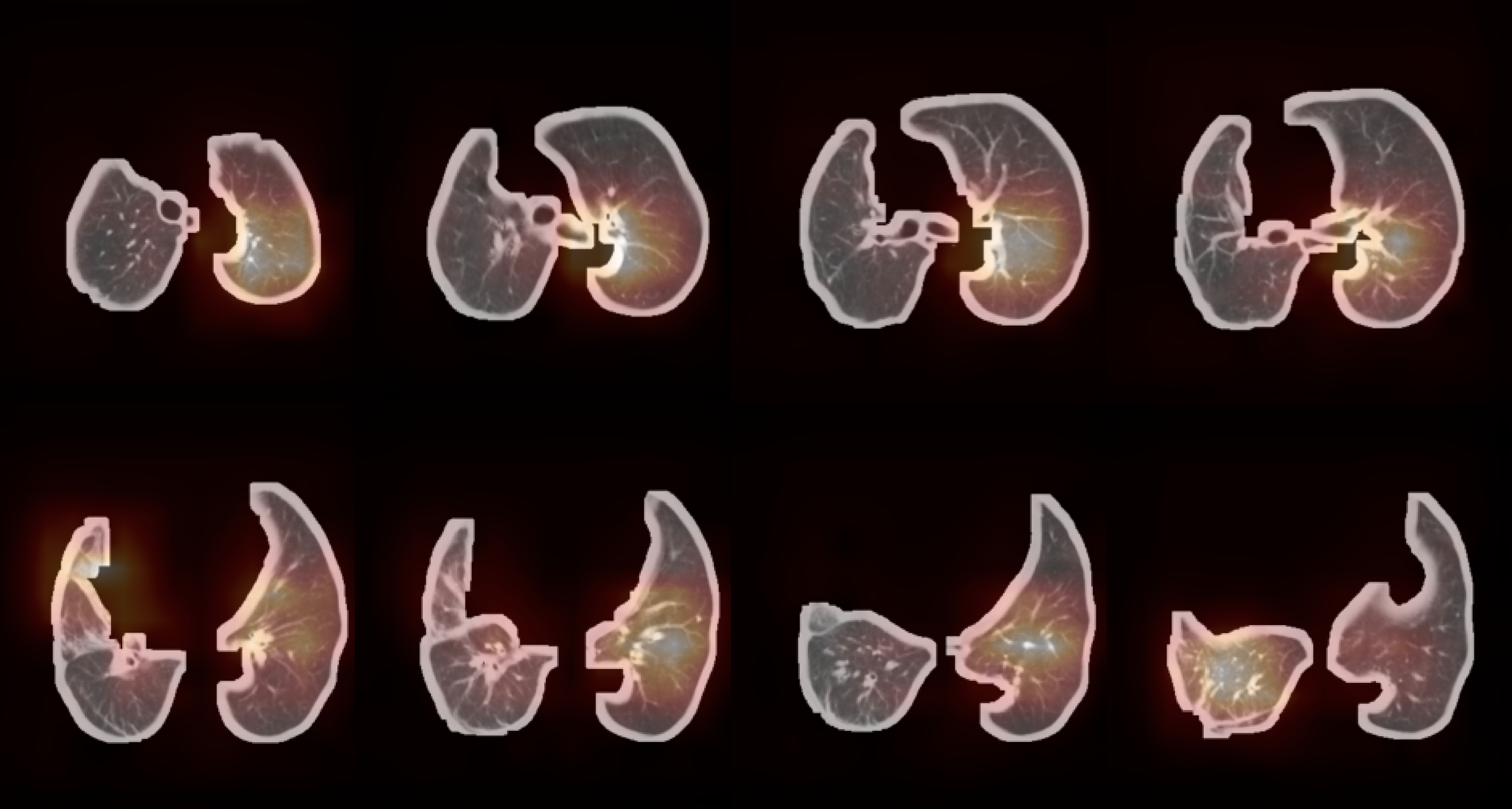}
    \caption{The class activation mappings generated on CovidNet3D on a chest CT scan of the Clean-CC-CCII dataset. Regions colored in red and brighter has more impact on model's decision to the class of COVID-19 while blue and darker region has less.}
    \label{fig:cam}
\end{figure}

\subsection{Interpretability}
Although our model achieves promising result in detecting COVID-19 in CT images, classification result itself does not help clinical diagnosis without proving the inner mechanism which leads to the final decision makes sense. To inspect our CovidNet3D model's inner mechanism, we apply Class Activation Mapping (CAM) \cite{cam} on it.

CAM is an algorithm that can visualize the discriminative lesion regions that the model focuses on. In Fig. \ref{fig:cam}, we apply CAM on each slice of a whole 3D CT scan volume from Clean-CC-CCII dataset. Regions appear red and brighter have a larger impact on the model's decision to classify it to COVID-19. From the perspective of the scan volume, we can see that some slices have more impacts on the model's decision than the others. In terms of a single slice, the areas that CovidNet3D focuses on has ground-glass opacity, which is proved a distinctive feature of CT images of COVID-19 Chest CT images \cite{radio-chestCT}. CAM enables the interpretability of our searched models (CovidNet3D), helping doctors quickly locate the discriminative lesion areas.






\section{Conclusion}

In this work, we introduce the differentiable neural architecture (DNAS) framework combined with the Gumbel Softmax technique to search for 3D models on three open-source COVID-19 CT scan datasets. The results show that CovidNet3D, a family of models discovered by DNAS can achieve comparable results to the baseline 3D models with smaller size, which demonstrates that NAS is a powerful tool for assisting in COVID-19 detection. In the future, we will apply NAS to the task of 3D medical image segmentation to locate the lesion areas in a more fine-grained manner. 

\section{Acknowledgments}

The research was supported by the grant {RMGS2019\_1\_23} from Hong Kong Research Matching Grant Scheme . We would like to thank the anonymous reviewers for their valuable comments.



\bibliographystyle{aaai}
\bibliography{main}

\end{document}